\documentclass{PoS}
\usepackage{natbib}

\title{Radio and $\gamma$-ray connection in relativistic jets}

\ShortTitle{Radio and $\gamma$-ray connection in relativistic jets}

\author{\speaker{M. Orienti}$^{1}$\thanks{orienti@ira.inaf.it}\\
$^1$INAF -- IRA, Bologna, Italy \\
}

\abstract{Relativistic jets are one of the most powerful manifestations 
of the
release of energy related to the supermassive black holes at the
centre of active galactic nuclei (AGN).
Their emission is observed across the entire
electromagnetic spectrum, from the radio band to gamma rays. Despite
decades of efforts, many aspects of the physics of relativistic jets
remain elusive. In particular, the location and the mechanisms
responsible for the high-energy emission and the connection of the
variability at different wavelengths are among the greatest challenges
in the study of AGN. Recent high resolution radio observations of 
flaring objects locate the high-energy emitting region downstream the 
jet at parsec scale distance from the central engine, posing questions
on the nature of the seed photons upscattered to $\gamma$-rays. Furthermore, 
monitoring campaigns of the most active blazars indicate that not all 
the high energy flares have the same characteristics in the various 
energy bands, even from the same source, making the interpretation of 
the mechanism responsible for the high-energy emission not
trivial. Although the variability of the most luminous blazars is
well explained 
by the ``shock-in-jet'' scenario, the sub-class of TeV emitting objects
suggests a more complex emission model with velocity gradients in a
structured jet.  
This contribution presents results obtained by recent multiwavelength 
campaigns of blazars aimed at studying the radio and
$\gamma$-ray connection and the physical mechanisms at the basis of
the emission in these low and high energy bands. 
}

\FullConference{12th European VLBI Network Symposium and Users Meeting\\
		7-10 October 2014\\
		Cagliari, Italy}

\begin{document}

\section{Introduction}

\noindent The $\gamma$-ray sky is dominated by the population of active galactic
nuclei (AGN). AGN are about 58\% of the objects in the third
{\it Fermi} Large Area Telescope source catalog (3FGL; Acero et al. 2015), 
which reports the
$\gamma$-ray sources detected by {\it Fermi}-LAT in the first four
years of scientific observations. A part from cosmic-ray related high
energy emission from 
a handful of normal galaxies \citep[e.g.,][]{ackermann12}, almost all the
extragalactic sources are associated with radio-loud AGN. 
In radio-loud AGN the radio emission is comparable or even stronger
than the emission observed in the other energy bands, and is
associated with the presence of a bipolar outflow of relativistic
plasma that is responsible for both
synchrotron (from radio up to X-rays) and inverse Compton radiation
(at high energies).\\ 
The large majority ($\sim$98\%) of the AGN emitting at high energies
is represented by the blazar population which comprises flat spectrum
radio quasars (FSRQ) and BL Lac objects. In these objects the
relativistic jet is aligned to our line of sight and the emission is amplified
by Doppler boosting effects. The remaining 2\% are associated
with misaligned AGN, as reported in the third catalog of AGN detected
by {\it Fermi}-LAT  \citep[3LAC;][]{ackermann15}. It is worth
noting that 
during the EGRET era, only three radio galaxies were
proposed as the low-energy counterpart of $\gamma$-ray sources
\citep[Centaurus\,A, NGC\,6251, J1737-15;][]{hartman99}, but
\citet{casandjian08} did not 
confirm the association. \\
The dominance of the radio-loud population among $\gamma$-ray emitting
objects have suggested a possible relation between the emission in
radio and $\gamma$-ray energy bands. This connection is supported by
the spectral energy distribution (SED) of the blazar sub-classes, which are
well-modeled just by changing the radio luminosity. This suggests
common under-lying physical processes linking the synchrotron and the
inverse Compton emission \citep{fossati98}.\\ 
Although $\gamma$-ray-loud AGN
are radio-loud 
sources, this does not imply that all the radio-loud AGN are
$\gamma$-ray emitters. In particular, blazars with strong $\gamma$-ray
emission are brighter and more luminous at radio frequencies 
\citep{kovalev09}, have faster jets \citep{lister09} and higher
variability Doppler factors \citep{savolainen10} with respect to
those without (significant) $\gamma$-ray emission, indicating that
{\it Fermi}-LAT mainly detects the brightest objects from
radio flux-density-limited samples.\\
Despite decades of efforts, many aspects of the physics of
relativistic jets remain elusive. In particular, understanding how the jet is
launched and accelerated, what is its structure, where is the
high-energy emitting region and what is the responsible mechanism, are
among the greatest challenges in the study of AGN. So far, many
scenarios have been proposed for describing the typical variability
observed in blazars. The ``shock-in-jet'' model predicts that a
disturbance in the jet forms a shock that propagates downstream along
the jet and is responsible for both the high and low energy emission
\citep[e.g.,][]{marscher85}. On the other hand, the ``structured-jet
scenario'' predicts that the jet has a velocity structure, and the
high- and low-energy emission comes from different regions
\citep[e.g.,][]{ghisellini05}.\\
%
%
Understanding what
drives the emission at low and high energies will provide strong
constraints on the open questions about the physics of the
relativistic jets. By combining information arising from (high
resolution) radio observations and $\gamma$-ray light curves it will
be possible to investigate the presence of shocks, turbulence, or
velocity gradients in the jet, the nature of the seed photons upscattered
at high energy, as well as the magnetic field structure.\\
This contribution reports on some recent 
results on the radio/$\gamma$-ray connection and on the
physical mechanisms that may relate the emission in these two bands.\\

\section{Radio/GeV connection}

\subsection{Statistical studies}

Many statistical works have been carried out in order to confirm, or
not, a radio/$\gamma$-rays connection. A detailed statistical analysis 
of the correlation between radio and $\gamma$-ray emission of AGN
detected by {\it Fermi} during its first year of operation was
presented in \citet{ackermann11}. This work made use of the
largest data set ever used so far. They
found that the statistical significance of a positive correlation
between the centimeter radio emission and the broadband (100 MeV $<$E
$<$ 300 GeV) 
$\gamma$-ray energy flux is very high for the whole AGN sample, and
increases when FSRQ and BL Lacs are considered separately. A similar
result was suggested by \citet{ghirlanda10} who considered a much smaller
sample. The radio-$\gamma$-ray correlation becomes very weak when energies E$>$10 GeV are considered, suggesting a physical rather than an
observational origin for explaining such a lack of correlation
\citep{giroletti15}. \\
The coexistence of {\it Fermi} and {\it Planck} satellites in orbit
has enabled the exploration of the connection between the
$\gamma$-rays and the radio emission at (sub-)millimeter
wavelengths. In their work, \citet{leon12} found a correlation between
the $\gamma$-ray and (sub-)mm luminosity which holds over five
orders of magnitude. However, in some bands the correlation 
becomes more significant when quasi-simultaneous observations (within
2 months) are considered. Observations of some time delay between the
variability in the different energy bands have implication on the
mechanism at the basis of the flare and its location. 
For this
reason, important information on the physics of relativistic jets may be
obtained by exploring the multi-band variability emission. \\

\subsection{Radio/GeV correlated variability}

Blazars show strong variability across the entire electromagnetic spectrum and
a simultaneous or delayed occurrence may indicate the location
of the variability region. Delayed variability at the long wavelengths
is expected in presence of opacity effects from the innermost compact region of
the AGN (within the broad line region, BLR). On the other hand, a
rough simultaneity of radio and GeV flares locates the production of
the $\gamma$-ray photons in a 
region downstream the jet, where the radio emission is not self-absorbed.\\
Because {\it Fermi} operates in an all-sky scanning mode, it is possible to
compare the $\gamma$-ray light curves with radio light curves from dedicated
monitoring programmes. \citet{fuhrmann14} presented results of a
cross-correlation analysis between $\gamma$-rays and cm/mm wavelengths
of a sub-sample of {\it Fermi}-bright blazars. They found a
highly significant correlation between the two bands, with the radio
lagging $\gamma$-rays. The time delay between high energies and radio
is (7$\pm$9) days at 142 GHz and
systematically increases as longer wavelengths are
considered, becoming (76$\pm$23) days at 2 GHz. 
The frequency-dependent time lag is in agreement with the
opacity of the jet. This suggests that the $\gamma$-rays
instantaneously escape from the production site, while the radio
emission from the same region becomes optically thin progressively at
later times, as a consequence of energy losses. The short time delay
between $\gamma$-rays and (sub-)millimeter emission indicates that
the $\gamma$-ray site is located about 1 pc upstream the mm core region.\\
A different conclusion was drawn by \citet{leon11} who compared 37-GHz
and $\gamma$-ray light curves. They noticed that statistically the
rising of the mm 
light curve precedes the $\gamma$-ray peak of a few months
(observer's frame),
suggesting that the high energy flare takes place a few parsecs
downstream along the jet. This opens questions on the seed photons
which are upscattered to the high energies. In fact, at such a
distance from the central region, the UV photons from the BLR are not
effective anymore, and other seed photons, like the IR from the dusty
torus \citep[e.g.][]{sikora08}, or synchrotron photons from a
different electron population must be considered. 
On the other hand, the strongest $\gamma$-ray flares occur during the
rising/peaking stages 
of the mm light curve, suggesting that both episodes are related to
the same disturbance. \\

\section{The role of VLBI}

The studies described in the previous Sections refer mainly to radio
observations with single-dish telescopes. Low resolution observations 
cannot separate the contributions from the various source components.
Observations with parsec-scale resolution are required to disentangle the
flux density and the polarization of the core region from the emission arising 
from the jet and extended features. \\
Important information on the innermost region of relativistic jets comes from
dedicated studies of bright blazars using the Very Long Baseline
Interferometry (VLBI) technique. The (sub-)milliarcsecond-scale
resolution of VLBI observations allowed a deep look into the jet base, 
revealing different structures, like the presence of superluminal knots
\citep[e.g.,][]{jorstad01a}, changes in the jet direction,
\citep[e.g., OJ\,287;][]{agudo11}, limb-brightened
structures \citep[e.g., M\,87;][]{kovalev07}.\\
Monitoring campaigns of multi-frequency VLBI observations of
$\gamma$-ray blazars detected by EGRET suggest that the highest level of
$\gamma$-ray emission is connected 
to the ejection of a new superluminal component
\citep[e.g.,][]{jorstad01b}. This supports the idea that at least the 
strongest $\gamma$-ray emission is strictly related to a shock in the
jet. \\ 

\subsection{Shock-in-jet model}

In the shock-in-jet model, the flare originates as a
disturbance that 
modifies the flow parameters and produces a shock wave
\citep{marscher85, valtaoja92}. The shock
model implies: (1) a growth stage, when the shock forms up to the
development of its maximum, which is observed not simultaneously at
the various energy bands due to opacity effects (Compton losses
dominate); (2) a plateau, when energy losses and gains are balanced
(synchrotron losses dominate); (3) a decaying stage, when the shock
fades due to energy losses (adiabatic losses
dominate). \\
The propagation of a
transverse shock produces 
the amplification of the perpendicular component of the magnetic field
with respect to the parallel one, and an enhancement of the
luminosity. However, if the shock is
oblique the polarization 
properties may be different, depending on the obliqueness of the shock
itself and to the characteristics of the underlying magnetic field
like its order and strength \citep{hughes11}. It is worth noting that
90$^{\circ}$ rotation of the electric vector position angle (EVPA) are
expected during the transition 
between the opacity regimes. However, such changes in the opacity would cause a
dramatic drop of the flux density which is not observed.\\    
%
%
%
The shock-in-jet scenario naturally explains the frequency-dependent
variability observed in many blazars, the detection of superluminal
components that
are interpreted as the observable manifestation of the propagating
shock, and the abrupt rotation of 90$^{\circ}$ of the EVPA \citep[e.g.,][]{orienti13,orienti11,hovatta08}.  
Outstanding examples are the FSRQ 3C\,454.3 and
PKS\,1510-089. These objects underwent major flares in $\gamma$-rays,
reaching an apparent peak luminosity $\sim$10$^{49-50}$ erg s$^{-1}$
\citep{orienti13,abdo11}. The huge flares occurred near the peak of
the millimeter flare and were accompanied by the
emergence of a superluminal knot with apparent speed above 10-20$c$,
indicating a tight connection between $\gamma$-ray flares and changes in the
parsec-scale jet structure \citep{jorstad13,orienti13}. However, the
location of the $\gamma$-ray production region is not unambiguously
determined. 
The short time lag derived between the $\gamma$-ray flare
and the detection of both a new component and the mm flare indicates
that the high energy region is located in the millimeter photosphere,
i.e. the region where the mm emission becomes optically thin, a few
parsecs away from the BLR. \\
In OJ\,287 \citet{agudo11} reported that
$\gamma$-ray peaks are likely 
associated with a large increase of the linear polarization from a
jet component located about 14 parsec from the central engine, strongly
indicating that the dissipation region is located well downstream
along the jet as a consequence of the interaction between a
disturbance and a standing conical shock.
If we consider that the entire cross-section of the jet is responsible
for the $\gamma$-ray emission, this scenario is rather difficult to
reconcile with the size expected on the basis of the causality
argument, unless the 
collimation angle of the jet is 
extremely small, or the flaring region occupies only a fraction of the
jet cross section. The latter scenario is supported by the
observations of frequent changes and large jumps in the position angle of the
direction of the superluminal knots moving along the jet
\citep{lister13}.\\
In PKS\,1510-089 no correlation between
the jet properties and the $\gamma$-ray flares was found, indicating
that the high-energy emission does not arise from the knot itself, but it
is likely located in the unresolved core component. By following the
evolution of three superluminal knots in PKS\,1510-089,
\citet{orienti13b} found that the flux density decreases with time as
expected for adiabatic losses, and the polarization percentage increases
reaching values up to 10\%, much larger than what is
found in the core component. Remarkably, all the knots
are characterized by the same EVPA $\sim$80$^{\circ}$, once they
emerge from the core.\\
Not all the $\gamma$-ray flares have a clear counterpart at low
energies even if they are produced by the same source, like in the case of
some outbursts of 3C 279 and BL 
Lac where no
significant radio variability was detected after the $\gamma$-ray flare
\citep{abdo10c,marscher08}. 
In this case the trigger should be a
disturbance that develops in the innermost region of the AGN, within
the BLR, where the emission is utterly opaque at the radio
wavelengths. \\
It is worth mentioning that not all the radio flares and the ejection
of new superluminal components are associated with high energy
flares. Straightforward examples are the Narrow Line Seyfert 1
SBS\,0846+513 and the FSRQ 
4C\,49.22 where the radio peak and the ejection of a new component are
not related to any obvious high energy activity
\citep{dammando13,cutini14}. The lack of a high energy counterpart may
be related to the absence of enough seed photons in the region where the
energy dissipation takes place. 
On the other hand, in these objects high
$\gamma$-ray activities occur during low radio activity states,
suggesting different regions and/or mechanisms for low and high energy
variability.\\

\subsection{The Structured-jet model}

Although it proved to be successful in describing many aspects of the
most luminous blazars, the shock-in-jet model seems inadequate for
reproducing the observational properties and the spectral energy
distribution of TeV blazars and radio galaxies. Although the high
energy emission implies high Doppler factors and extreme bulk motion
VLBI observations of TeV objects did not reveal any superluminal jet
component. \\ 
The properties of TeV objects are better
explained by the structured-jet scenario, which postulates a velocity
gradient in the jet, either transverse (faster central spine
surrounded by a slower layer; e.g. Ghisellini et al. 2005) or radial (the flow 
decelerates as it moves outward; e.g. Georganopoulos \& Kazanas
2003). The spine-layer model is supported by observations in TeV
objects of
limb-brightened structure both in total intensity 
(e.g., the BL Lac Mrk\,501 and the radio galaxy M87; Giroletti et
al. 2008, Kovalev et al. 2007) and in polarized emission
\citep[e.g., the BL Lac Mrk\,421;][]{lico14}. In this scenario the
electrons of each region upscatter the beamed soft 
photons coming from the other. \\
The lack of a unique zone responsible
for high and low energy emission may produce a large variety of
correlation, as well as uncorrelated variability. This seems the case
of the radio galaxy 3C\,84 at the centre of Perseus cluster. The
$\gamma$-ray and radio light curves of 3C\,84 do not seem correlated,
and no obvious radio flare is associated with the episodes of Very
High Energy (VHE, E$>$100 GeV) emission
\citep{nagai12}. Interestingly, this source was not detected 
in $\gamma$-rays
by EGRET. After three months of scientific observations, {\it
  Fermi}-LAT detected 
$\gamma$-ray emission from 3C\,84 at a
level 7 times higher than the EGRET upper limit, indicating that the
lack of $\gamma$-ray emission during the EGRET era was intrinsic to the
source rather 
than due to sensitivity limitation \citep{abdo09c}. A change in the physical
characteristics of 3C\,84 is marked by the different structure of the
jet: edge-darkened in the EGRET era, and limb-brightened in the {\it Fermi} era
\citep{nagai14}. This poses strong evidence in favour of the
jet-structured scenario for this misaligned AGN.\\

\section{Concluding remarks}

Multiwavelength monitoring campaigns suggest a relation
between $\gamma$-ray flares and the radio variability. However, the
time delay is different if the peak or the rising of the mm outburst
are considered. In the former case the $\gamma$-ray peak precedes the
radio one with a time lag increasing as longer wavelengths are
considered. On the other hand, in case of strong $\gamma$-ray flares,
the $\gamma$-ray flare seems to follow the onset of the radio outburst
suggesting that the dissipation region occurs at parsec-scale
distance from the central engine. The observed variability is usually 
explained in terms of a shock moving along the jet, whose
manifestation is a superluminal knot observable with high-frequency
VLBI observations. However, the variability
observed in $\gamma$-rays does not always show the same
properties in the other energy bands, and it is possible that
different $\gamma$-ray flares of the same source originate at
different distances from the central engine.\\
The ``shock-in-jet'' model is not able to reproduce the observational
properties of low-luminosity blazars. For example, the
$\gamma$-ray properties of the TeV
emitting blazars and radio galaxies are better explained by a two-zone
model that assumes a gradient velocity in the jet, either radial or
transverse. This
``structured-jet scenario'' is supported by observations of
limb-brightened structures of the jet, likely witnessing a faster
central spine surrounded by a slower layer.
Therefore, low and high energy emission originates in different regions
and a connection between the radio and $\gamma$-rays may not be trivial. \\
The advent of the Cherenkov Telescope Array, operating in the GeV/TeV
band, will allow us to investigate the physics of relativistic jets in
the most extreme objects which are not well sampled by the {\it
  Fermi}-LAT energy bands.\\


\begin{thebibliography}{99}



\bibitem[Abdo et al.(2009)]{abdo09c}
Abdo, A.A., Ackermann, M., Ajello, M., et al. 2009, ApJ, 699, 31



\bibitem[Abdo et al.(2010)]{abdo10c}
Abdo, A.A., Ackermann, M., Ajello, M., et al. 2010, Nature, 463, 919

\bibitem[Abdo et al.(2011)]{abdo11}
Abdo, A.A., Ackermann, M., Ajello, M., et al. 2011, ApJ, 733, 26

\bibitem[Acero et al.(2015)]{acero15}
Acero, F., Ackermann, M., Ajello, M., et al. 2015, ApJS submitted
(arXiv:1501.02003) 

\bibitem[Ackermann et al.(2011)]{ackermann11}
Ackermann, M., Ajello, M., Allafort, A., et al. 2011, ApJ, 741, 30

\bibitem[Ackermann et al.(2012)]{ackermann12}
Ackermann, M., Ajello, M., Allafort, A., et al. 2012, ApJ, 755, 164

\bibitem[Ackermann et al.(2015)]{ackermann15}
Ackermann, M., Ajello, M., Atwood, W.B., et al. 2015, ApJS submitted
(arXiv:1501:06054) 

\bibitem[Agudo et al.(2011)]{agudo11}
Agudo, I., Jorstad, S.G., Marscher, A.P., et al. 2011, ApJ, 726, 13


\bibitem[Casandjian \& Grenier(2008)]{casandjian08}
Casandjian, J.-M., Grenier, I.A. 2008, A\&A, 489, 849

\bibitem[Cutini et al.(2014)]{cutini14}
Cutini, S., Ciprini, S., Orienti, M., et al. 2014, MNRAS, 445, 4316

\bibitem[D'Ammando et al.(2013)]{dammando13}
D'Ammando, F., Orienti, M., Finke, J., et al. 2013, MNRAS, 436, 191

\bibitem[Fossati et al.(1998)]{fossati98}
Fossati, G., Maraschi, L., Celotti, A., Comastri, A., Ghisellini,
G. 1998, MNRAS, 299, 433 

\bibitem[Fuhrmann et al.(2014)]{fuhrmann14}
Fuhrmann, L., Larsson, S., Chiang, J., et al. 2014, MNRAS, 441, 1899

\bibitem[Georganopoulos \& Kazanas(2003)]{gk03}
Georganopoulos, M., Kazanas, D. 2003, ApJ, 594, 27

\bibitem[Ghirlanda et al.(2010)]{ghirlanda10}
Ghirlanda, G., Ghisellini, G., Tavecchio, F., Foschini, L. 2010,
MNRAS, 407, 791

\bibitem[Ghisellini et al.(2005)]{ghisellini05}
Ghisellini, G., Tavecchio, F., Chiaberge, M. 2005, A\&A, 432, 401

\bibitem[Giroletti et al.(2008)]{giroletti08}
Giroletti, M., Giovannini, G., Cotton, W.D., et al. 2008, A\&A, 488, 905

\bibitem[Giroletti et al.(2015)]{giroletti15}
Giroletti, M., Lico, R., Bignall, H., et al. 2015, PoS(EVN2014)014

\bibitem[Hartman et al.(1999)]{hartman99}
Hartman, R.C., Bertsch, D.L., Bloom, S.D., et al. 1999, ApJS, 123, 79

\bibitem[Hovatta et al.(2008)]{hovatta08}
Hovatta, T., Nieppola, E., Tornikoski, M., et al. 2008, A\&A, 485, 51

\bibitem[Hughes et al.(2011)]{hughes11}
Hughes P.A., Aller M.F., Aller H.D. 2011, ApJ, 735, 81

\bibitem[Jorstad et al.(2001a)]{jorstad01a}
Jorstad, S.G., Marscher, A.P., Mattox, J.R., et al. 2001a, ApJS, 134,
181

\bibitem[Jorstad et al.(2001b)]{jorstad01b}
Jorstad, S.G., Marscher, A.P., Mattox, J.R., et al. 2001b, ApJ, 556, 738

\bibitem[Jorstad et al.(2013)]{jorstad13}
Jorstad, S.G., Marscher, A.P., Smith, P.S., et al. 2013, ApJ, 773, 147

\bibitem[Kovalev et al.(2007)]{kovalev07}
Kovalev, Y.Y., Lister, M.L., Homan, D.C., Kellermann, K.I. 2007, ApJ,
668, 27

\bibitem[Kovalev et al.(2009)]{kovalev09}
Kovalev, Y.Y., Aller, H.D., Aller, M.F., et al. 2009, ApJ, 696, 17

\bibitem[Leon-Tavares et al.(2011)]{leon11}
Le\'on-Tavares, J., Valtaoja, E., Tornikoski, M., L\"ahteenm\"aki, A.,
Nieppola, E. 2011, A\&A, 532, 146

\bibitem[Leon-Tavares et al.(2012)]{leon12}
Le\'on-Tavares, J., Valtaoja, E., Giommi, P., et al. 2012, ApJ, 754, 23

\bibitem[Lico et al.(2014)]{lico14}
Lico, R., Giroletti, M., Orienti, M., et al. 2014, A\&A, 571, 54

\bibitem[Lister et al.(2009)]{lister09}
Lister M.L., Homan D.C., Kadler M., et al. 2009, ApJ, 696, 22

\bibitem[Lister et al.(2013)]{lister13}
Lister, M.L., Aller, M.F., Aller, H.D., et al. 2013, AJ, 146, 120


\bibitem[Marscher \& Gear(1985)]{marscher85}
Marscher A.P., Gear W.K. 1985, ApJ, 298, 114

\bibitem[Marscher et al.(2008)]{marscher08}
Marscher, A.P., Jorstad, S.G., D'Arcangelo, F.D., et
al. 2008, Nature, 452, 966


\bibitem[Nagai et al.(2012)]{nagai12}
Nagai, H., Orienti, M., Kino, M., et al. 2012, MNRAS, 423, 122

\bibitem[Nagai et al.(2014)]{nagai14}
Nagai, H., Haga, T., Giovannini, G., et al. 2014, ApJ, 785, 53

\bibitem[Orienti et al.(2011)]{orienti11}
Orienti, M., Venturi, T., Dallacasa, D., et al. 2011, MNRAS, 417, 359

\bibitem[Orienti et al.(2013a)]{orienti13}
Orienti, M., Koyama, S., D'Ammando, F., et al. 2013a, MNRAS, 428, 2418

\bibitem[Orienti et al.(2013b)]{orienti13b}
Orienti, M., D'Ammando, F., Giroletti, M., et al. 2013b, EPJWC, 6104009

\bibitem[Savolainen et al.(2010)]{savolainen10}
Savolainen T., Homan D.C., Hovatta T., et al. 2010, A\&A, 512, 24

\bibitem[Sikora et al.(2008)]{sikora08}
Sikora, M., Moderski, R., Madejski, G.M. 2008, ApJ, 675, 71

\bibitem[Valtaoja et al.(1992)]{valtaoja92}
Valtaoja, E., Terasranta, H., Urpo, S., et al. 1992, A\&A, 254, 71

\end{thebibliography}
\end{document}